\documentclass[a4paper,twocolumn,prl,superscriptaddress,amsmath,amssymb,mathrsfs]{revtex4}

\usepackage{hyperref}
\hypersetup{%
    colorlinks,
    linkcolor={black},
    citecolor={blue},
    urlcolor={black}}

\usepackage{braket}
\usepackage{epstopdf}
\usepackage{graphicx}
\usepackage{mathtools} 
\usepackage{dsfont} 
\usepackage{ifthen}
\usepackage{units}
\usepackage{bm}
\usepackage[percent]{overpic}
\usepackage{upgreek} 
\usepackage{titlesec}
\usepackage[titletoc,title]{appendix}

\newcommand{\ssec}[1]{\emph{#1} ---}
\newcommand{\figref}[2][]{Fig.~\ref{#2}\ifthenelse{\equal{#1}{}}{}{(#1)}}
\newcommand{\Figref}[2][]{Figure~\ref{#2}\ifthenelse{\equal{#1}{}}{}{(#1)}}
\newcommand{\myeqref}[1]{Eq.~(\ref{#1})}

\newcommand{\op}[2][]{\hat{#2}\ifthenelse{\equal{#1}{}}{}{_\text{#1}}} 
\newcommand{\Op}[2][]{\hat{\bm{#2}}\ifthenelse{\equal{#1}{}}{}{_\text{#1}}} 
\newcommand{\res}[2]{\ensuremath{\ket{\downarrow,m_\ell=#1;\nu^-,M}
                                 \leftrightarrow
                                 \ket{\uparrow,m_\ell=#2;\nu^+,M'}}}
\newcommand{\initprodstate}[2]{\ensuremath{\ket{\downarrow,m_\ell=#1}\otimes\ket{\nu^-,M=#2}}}
\newcommand{\finlprodstate}[2]{\ensuremath{\ket{\uparrow,  m_\ell=#1}\otimes\ket{\nu^+,M=#2}}}

\newcommand{\potdd}{\ensuremath{\op[dd]{V}}}
\newcommand{\coup}{\ensuremath{V_\text{dd}}}
\newcommand{\rmol}{\ensuremath{\bm{r}_\text{mol}}} 
\newcommand{\rmola}{\ensuremath{r_\text{mol}}} 
\newcommand{\ptrans}{\ensuremath{P_\text{ex}}} 

\newcommand{\kHz}[1]{\unit[#1]{kHz}}

\newcommand{\mps}[1]{\unit[#1]{m/s}}
\newcommand{\cmtwo}[1]{\unit[#1]{cm^2}}

\newcommand{\nm}[1]{\unit[#1]{nm}}
\newcommand{\Vpm}[1]{\unit[#1]{V/m}}
\newcommand{\microsec}[1]{\unit[#1]{\upmu s}}
\newcommand{\microm}[1]{\unit[#1]{\upmu m}}

\begin{document}

\title{Rydberg atom-enabled spectroscopy of polar molecules via F\"orster resonance
energy transfer}

\author{Sabrina Patsch}
\affiliation{%
  Dahlem Center for Complex Quantum Systems and Fachbereich Physik,
  Freie Universit\"{a}t Berlin, Arnimallee 14, 14195 Berlin, Germany
}

\author{Martin Zeppenfeld}
\affiliation{%
  Max-Planck-Institut f\"{u}r Quantenoptik, Hans-Kopfermann-Stra\ss e 1, 85748
  Garching, Germany
}

\author{Christiane P. Koch}
\affiliation{%
  Dahlem Center for Complex Quantum Systems and Fachbereich Physik,
  Freie Universit\"{a}t Berlin, Arnimallee 14, 14195 Berlin, Germany
}
\email{christiane.koch@fu-berlin.de}

\date{\today}

\begin{abstract}
  Non-radiative energy transfer between a Rydberg atom and a polar molecule can
  be controlled by a DC electric field. Here we show how to exploit this control
  for state-resolved, non-destructive detection and spectroscopy of the
  molecules where the lineshape reflects the type of molecular transition. Using
  the example  of ammonia, we identify the conditions for collision-mediated
  spectroscopy in terms of the required electric field strengths, relative
  velocities, and molecular densities. Rydberg atom-enabled spectroscopy is
  feasible with current experimental technology, providing a versatile detection
  method as basic building block for applications of polar molecules in quantum
  technologies and chemical reaction studies.
\end{abstract}

\maketitle

Cold polar molecules are an excellent platform for quantum control with
applications ranging from fundamental physics~\cite{HutzlerQST20,MitraPRA2022}
and quantum information~\cite{AlbertPRX20,Wang2022,Zhang2022} to cold
chemistry~\cite{bookDulieuOsterwalder}. The ability to detect the molecules,
ideally at the single molecule level and in a non-destructive and state-resolved
fashion, is a prerequisite to any such application. Optical detection schemes
such as laser-induced fluorescence  or
absorption~\cite{Shuman2009,Wang2010,Cheuk2018,Shaw2021} are destructive and
difficult to apply at low density except for select molecules with optical
cycling transitions.  An alternative approach to non-destructive detection
suggested use of Rydberg atoms~\cite{Kuznetsova2016,Zeppenfeld2017} -- instead of driving
molecular transitions by  laser light, it exploits F\"orster resonant energy
transfer (FRET), i.e.,  non-radiative energy exchange between donor and
acceptor mediated by resonant dipole-dipole
interaction~\cite{Foerster1948,AndrewsBook1999}. Rydberg atoms are particularly
well suited to FRET due to their large dipole
moment~\cite{SafinyaPRL1981,RavetsNatPhys2014}. Rydberg states are readily
prepared~\cite{Haroche2006,Larrouy2020}, and the scaling of Rydberg transitions
with the principal quantum number~\cite{Sibalic2018} allows for covering the
microwave and terahertz spectral range, i.e., rotational transition frequencies
in a large variety of molecules~\cite{Zeppenfeld2017}. Basic feasibility of
non-destructive detection of molecules via FRET with Rydberg atoms has been
demonstrated for ammonia~\cite{Jarisch2018}, as has electric field control of
the energy transfer~\cite{Jarisch2018,ZhelyazkovaPRA2017,Gawlas2020}. The latter
leverages the easy tunability of Rydberg energy levels, due to their sensitivity
to external fields~\cite{Facon2016,Adams2020}. The tunability suggests that
FRET with a Rydberg atom may not only allow one to see whether a molecule is
present or not but to actually infer the molecular state prior to the
interaction. This would be an extremely useful tool for  quantum technologies
and studies of cold and ultracold chemistry but obviously requires a description
of the molecular structure beyond the popular two-level
approximation~\cite{Kuznetsova2018,Wang2022,Zhang2022,Zeppenfeld2017,Jarisch2018,ZhelyazkovaPRA2017,Gawlas2020}.

Here, we establish a theoretical framework from first principles for FRET in
collisions of polar molecules and Rydberg atoms and predict electric
field-dependent cross sections with full account of the interparticle dynamics.
These cross sections will be obtained in an experiment  by measuring the final
state of the Rydberg atom via, e.g., ionization.
The key parameter governing a collision is the relative velocity.
We find that at sufficiently low
relative velocity the cross sections display well-resolved lines as
a function of electric field, with intricate lineshapes that encode the relevant
selection rules. At very low velocity, linewidths below 1$\,$MHz are
achievable.  Analysis of the peak heights and positions allows for inferring the
state of the molecule or molecular ensemble.  Such Rydberg spectroscopy of polar
molecules only requires existence of (near) resonant dipole transitions in the
two particles.

  \begin{figure}
    \includegraphics{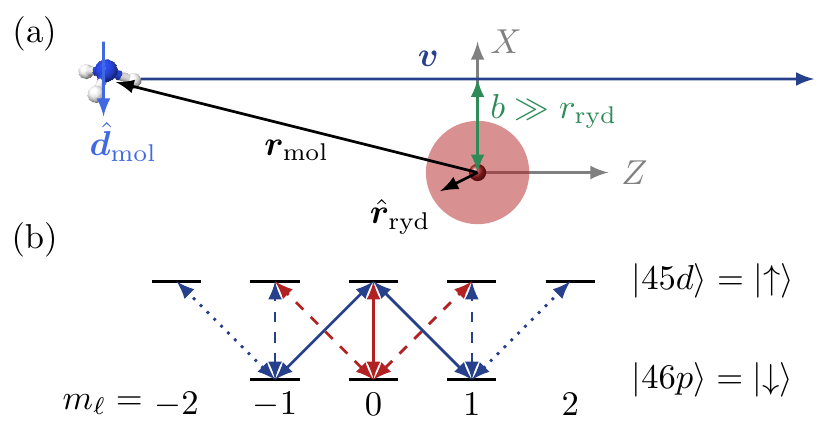}
    \caption{(Colour online)
      Collision-mediated Rydberg spectroscopy of polar molecules relies on the
      interaction of the molecular dipole moment (lightblue arrow) with the
      electric field due to the charge distribution of the Rydberg atom's
      valence electron (indicated by the shaded area).
      (a)~Classical scattering trajectory (dark blue).
      (b)~Relevant energy levels of the Rydberg atom with all possible dipole
      transitions indicated by arrows.
    }
    \label{fig1}
  \end{figure}

\ssec{Theoretical framework}
The different time and energy scales allow us to separate translational motion
(which we treat classically) and internal degrees of freedom~\cite{appendix}
similarly to the semi-classical impact parameter method~\cite{Beigman1995}.
The distance $\rmol(t)$ between molecule and atomic core, cf. \figref[a]{fig1},
is then a time-dependent parameter in the atom-molecule interaction.  The
Hamiltonian for the internal degrees of freedom is given by
\begin{equation}
  \op{H}(t)= \op[ryd]{H}
  + \op[ryd]{V}^\text{DC}
  + \op[mol]H
  + \op[int]{V}(t)\,,
  \label{eq:hamiltonian}
\end{equation}
where $\op[ryd]{V}^\text{DC}$ captures the Stark shifts in the Rydberg atom due
to the DC electric field~\footnote{The Stark effect of the molecule is
negligible at the relevant field strengths~\cite{Townes1975}.}.
$\op[int]{V}(t)$ describes the interaction of the molecular electric dipole
moment $\Op[mol]{d}$ with the electric field due to the charge distribution of
the Rydberg atom's valence electron. For sufficiently large distances between
atom and molecule, it is given by the dipole-dipole contribution,
\begin{equation}
  \potdd(t) =\frac{\Op[mol]{d}\cdot\Op[ryd]{r}}{\rmola^3(t)}
  -\frac{3\big( \Op[mol]{d} \cdot \rmol(t) \big)
    \big( \Op[ryd]{r} \cdot \rmol(t) \big)}
  {\rmola^5(t)}\,,
  \label{eq6:dip_pot}
\end{equation}
where $\Op[ryd]{r}$ is the position of the Rydberg electron.
We assume the ensemble to be sufficiently dilute such that a Rydberg
atom interacts at most once with a molecule~\cite{appendix}
and neglect any change of the molecular trajectory due to the interaction,
\begin{equation}
  \rmol(t) = v t \, \bm{e}_Z + b \, \bm{e}_X
  \label{eq:trajectory}
\end{equation}
with $v \bm{e}_Z$ the velocity and $b$ the impact parameter. Velocities down to
about $v=\mps{0.1}$ can be considered; for smaller velocities, the Rydberg atom
is likely to decay before the particles had enough time to interact~\cite{appendix}.
In \myeqref{eq:trajectory}, the molecular beam direction is parallel to the DC
field but extension to other orientations of the trajectory is straightforward.
The collision cross section for a given initial state $\ket{\Psi_0}$ of atom and
molecule is obtained by integrating the probability $\ptrans^{\Psi_0}(\Delta,b)$
for FRET to occur,
\begin{equation}
  \sigma_{\Psi_0}(\Delta) = 2\pi
  \int_0^\infty b \, db \, \ptrans^{\Psi_0}(\Delta,b)\,.
  \label{eq6:crosssec_phi0}
\end{equation}
Electric field tunability of the cross section arises from tuning the energy
mismatch $\Delta$ via the Stark effect of the Rydberg atom.

Motivated by recent experiments~\cite{Jarisch2018,Gawlas2020},
we consider FRET between Rydberg atoms and the inversion mode of ammonia.
The molecular state can be written in terms of the vibrational inversion mode
$\ket{\nu^\pm}$ and the symmetric top eigenstates $\ket{JKM}$~\cite{appendix}.
Since rotational transition frequencies are large compared to the inversion
splitting, it is sufficient to consider a single inversion doublet at a time,
such that $\ket{\nu^\pm JKM}$ with $\ket{\nu^+}$ the lowest, symmetric and
$\ket{\nu^-}$ the second lowest, anti-symmetric sub-level of the inversion
doublet for given $J$, $K$ and $M$. Coupling to other vibrational or electronic
degrees of freedom is negligible. The inversion splitting depends on $J$ and $K$
approximately as~\cite{Townes1975}
\begin{equation}
  \omega_\text{inv} = \omega_\text{inv}^0
  - c_1 \left( J(J+1) - K^2 + c_2 K^2\right)\,.
  \label{eq:w_inv}
\end{equation}
with constants $c_{1,2}$~\cite{Townes1975}.
For fixed $J$ and $K$, dipole transitions obey the selection rules $\nu^\pm
\leftrightarrow \nu^\mp$ and $\Delta M=\pm 1$ or $\Delta M = 0$ (unless
$M=0=M'$).

$\op H_\text{ryd}$ is represented in the spherical basis $\ket{n\ell,m_\ell}$
with principal quantum number $n$,  angular quantum number $\ell$ and projection
$m_\ell$. Assuming rubidium, levels with $\ell \leq 7$ are shifted by the
quantum defect $\delta_{n\ell j}$  due to the finite size of the ionic
core~\cite{Gallagher1994}. We neglect spin-orbit coupling such that
$\delta_{n\ell} \approx\delta_{n\ell,j=\ell+\frac{1}{2}}$. For low $J$ and
$K=J$, the inversion splitting of ammonia is matched by transitions between
$\ket{46p,m_\ell}\equiv\ket{\downarrow,m_\ell}$ and
$\ket{45d,m_\ell'}\equiv\ket{\uparrow,m_\ell'}$, cf. \figref[b]{fig1}.
For $J=K=1$, the energy mismatch between the atomic and molecular transitions
is shown in \figref[a]{fig2} as a function of the DC field strength. It
vanishes at different field strengths for transitions involving different
$m_\ell$. This is at the core of the suggested spectroscopy.


\begin{figure}
  \includegraphics{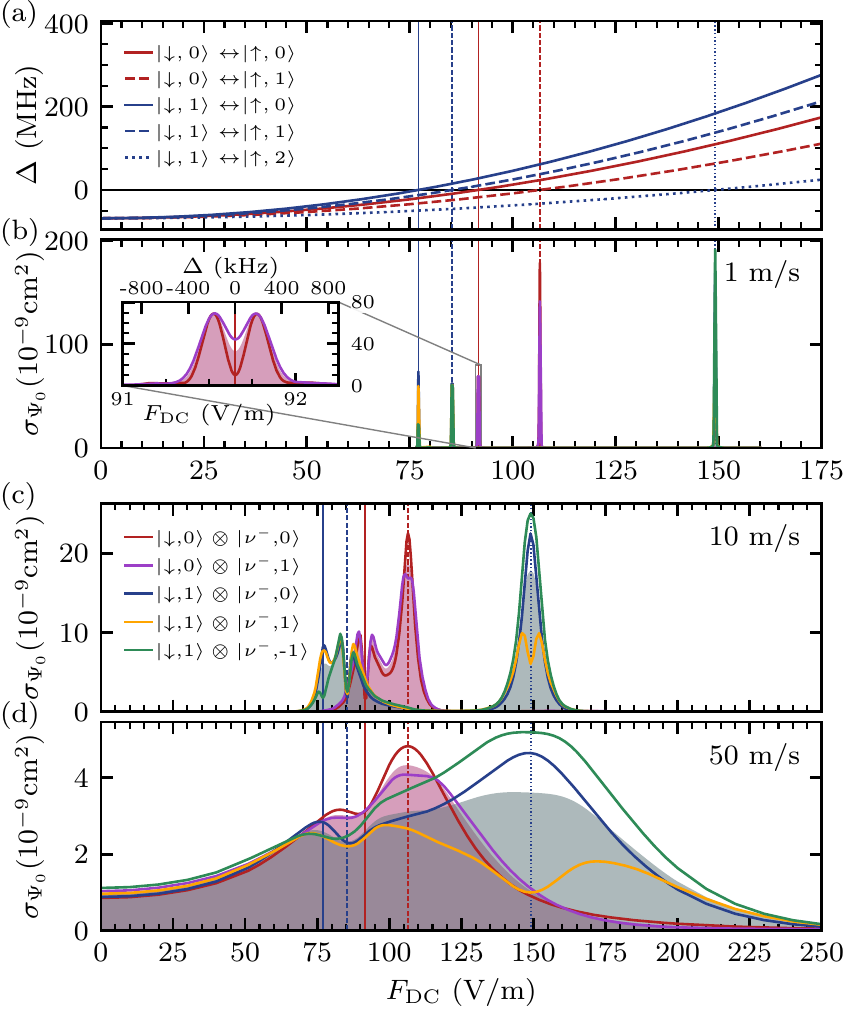}
  \caption{(Colour online)
    Electric field control of collisions between molecules prepared in
    a single rotational energy level (here $J=1=K$) and Rydberg atom:
    (a)~Energy mismatch $\Delta$ between the inversion mode of ammonia and the
    Rydberg transitions shown in \figref[b]{fig1}. Vertical lines indicate
    resonances.
    (b-d)~Cross sections for different initial states (indicated by line
    colour) and velocities (panels). Red (green) shaded areas show cross
    sections averaged over initial molecular states with the atom in
    $\ket{\downarrow,0}$ ($\ket{\downarrow,1}$).
    Note the different scales of the x- and y-axes.
  }
  \label{fig2}
\end{figure}

\ssec{Rydberg spectroscopy of polar molecules}
We consider two different scenarios for spectroscopy. Both
can be realized with existing experimental setups. In the first one, we assume
the molecule to be prepared in a single rotational energy level, for example
in an experiment with a suitable (e.g. quadrupole) guide or Stark
deccelerator~\cite{Petitjean1986Nat,Bethlem2002,Chervenkov2014}. In the
second scenario, we consider molecules in essentially arbitrary rotational
states taking a thermal state as example
and show that the cross sections allow for
inferring that state. The two scenarios differ in the number of
state-to-state cross sections that have to be averaged when predicting the
outcome of an experiment. Both rely on FRET between atom and molecule and both
assume state-resolved detection of the Rydberg atom after
the collision.


\ssec{Molecules in a single rotational level}
We assume $J=K=1$ for the molecule
and the Rydberg atom in one of the states $\ket{\downarrow,m_\ell}$ with
$m_\ell = 0, \pm 1$. Since the energy transfer is resonant, it is sufficient
to consider only the upper inversion level for the molecule,
$\ket{\nu^-,M}$ with $M=0,\pm 1$.
The cross section for all possible initial states is shown as a function of
the DC field strength for three relative velocities in \figref{fig2}(b-d)
with  resonances indicated by vertical lines.
As the velocity increases, the cross section peaks increase in width and decrease in amplitude. 
For dipole-dipole transitions,
the maximum of the cross section and its line width scale as~\cite{Zeppenfeld2017}
  \begin{align}
    \max{(\sigma)} \sim d_\text{mol}/v, \quad
    \Delta \sim v^{3/2}/\sqrt{d_\text{mol}}
    \label{eq:scaling}
  \end{align}
with $d_\text{mol} \sim K/\sqrt{J(J+1)}$ for the considered transition in
symmetric top molecules~\cite{Kroto1975}.  In a molecular beam experiment, the
molecules are randomly oriented, corresponding to an average over all $M$
states. In contrast, atomic states with different $m_\ell$ are easily selected
in the Rydberg state preparation.
We thus consider two $M$-averaged cross sections, for $m_\ell = 0$ and $m_\ell
= 1$ (shaded areas in \figref{fig2}(b-d)): At low velocities, the peaks do not
overlap (\figref[b]{fig2}) and  are clearly distinguishable also for higher
velocities (\figref[c]{fig2}).
While 
the resolution of single transitions is hampered at increasing velocity, the two
$M$-averaged cross sections continue to be distinguishable up until about
$v=50\,$m/s (\figref[c]{fig2}). In particular, the two resonances occurring at
the largest field strengths, corresponding to \res{0}{1} for the dashed red and
\res{1}{2} for the dotted blue vertical lines, can be resolved. It is thus
possible to deduce the inversion splitting and hence the rotational level from
the recorded cross-sections.  At higher velocities, this is no longer the case.

The peak heights in \figref[b-d]{fig2} imply that an effective volume of
$\unit[2\cdot 10^{-9}]{cm^3}$ is probed by one Rydberg atom, assuming an
interaction time of $\microsec{100}$. This suggests a fully saturated signal
at a molecular density of $\unit[5 \cdot 10^{8}]{cm^{-3}}$. In experiments,
the molecular signal has to be discriminated against the background, mainly
due to Rydberg transitions caused by blackbody radiation~\cite{Jarisch2018}.
Given the cross sections in \figref[b-d]{fig2}, this should be possible for
densities as low as $\unit[10^{5}]{cm^{-3}}$, cf. \cite{appendix} for details.

Various lineshapes are observed in \figref[b-c]{fig2}, even in the $M$-averaged
case. For example, if the Rydberg atom starts in $m_\ell = 0$ (vertical red
lines), the cross section around $\Vpm{107}$ (dashed) is Lorentzian but displays
a clear dip around $\Vpm{92}$ (solid, see inset). We now show that it is the
time-dependence of the interaction~\eqref{eq6:dip_pot} due to the collision that
is reflected in the lineshape.


\begin{figure}[tb]
  \includegraphics{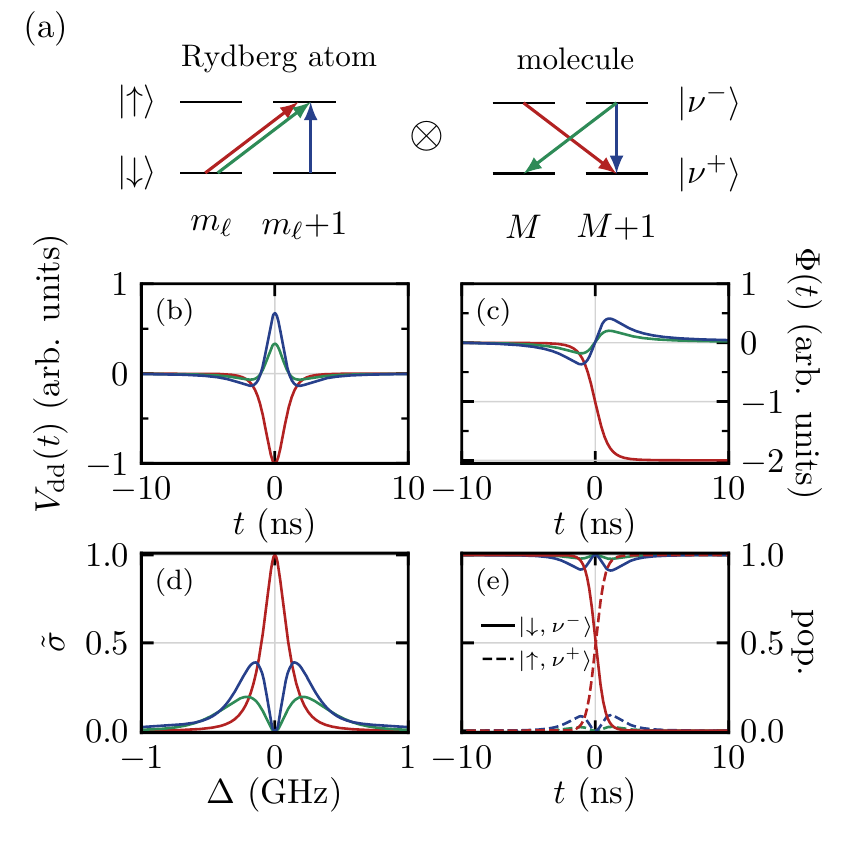}
  \caption{(Colour online)
    Types of transitions in two two-level systems due to dipole-dipole
    coupling (a) and their behaviour during the collision (b-e).
    (b)~interaction strength $\potdd(t)$ scaled to a maximal absolute value of
    $1$, (c)~its integrated value $\Phi(t)$;
    (e)~exemplary time evolution at resonance;
    (d)~normalized cross-section $\tilde{\sigma}= \sigma / \sigma_\text{max}$
    as a function of the energy mismatch $\Delta$ (with $\sigma_\text{max}
    = \cmtwo{1.7 \cdot 10^{-9}}$).
  }
  \label{fig3}
\end{figure}

\ssec{Understanding the lineshapes}
The selection rules allow for three types of FRET transitions, cf.
\figref[a]{fig3}, which we term criss-cross (red), and linear (blue) and
diagonal (green) flip-flop. In order to analyze each type separately, we reduce
the Hilbert space of atom and molecule to only two states each, the ones
connected by the respective transitions in \figref[a]{fig3}.  Then there is only
a single matrix element for the interaction, $\coup(t)
=\braket{\downarrow,m_\ell;\nu^-,M|\potdd(t)|\uparrow,m_\ell';\nu^+,M'}$, shown
in \figref[b]{fig3} for $v=100\,$m/s and $b=\nm{160}$. It is symmetric as
a function of time around $t=0$ where the two particles are closest to each
other and $\coup(t)$ takes its extremal value.  The two coupled two-level
systems (TLS) accumulate a relative phase, which in the resonant case is simply
given by $\Phi(t) = \int_{-\infty}^t \coup(t') dt'$, shown in \figref[c]{fig3}.
For the criss-cross transitions (red lines), $\coup(t)$ is always negative,
$\Phi(t)$ thus decreases monotonically, the two TLS exchange their excitation
perfectly (\figref[e]{fig3}), and
the TLS cross section displays a Lorentzian peak
(\figref[d]{fig3}). For flip-flop transitions (green and blue lines in
\figref[b-e]{fig3}), the non-Lorentzian lineshapes are rationalized by the
different time evolution of $\coup(t)$ with two changes of sign. Phase
accumulation is then non-monotonic and, most importantly, the phase equals zero
at $t=0$. This causes the TLS to return to their initial states, cf.\
\figref[e]{fig3}, resulting in a vanishing cross section at resonance.  Shifting
the  transitions away from resonance breaks the symmetry in the time evolution
of $\coup(t)$, resulting in non-vanishing accumulated phase and cross section.
This behavior is observed for both linear and diagonal flip-flop transitions
since only the overall strength of their coupling differs.

While the cross sections in \figref{fig2} arise from more complex
dynamics than that of two coupled TLS, the peaks resemble lineshapes
as seen in \figref[d]{fig3}. For example, the
peak at $\Vpm{107}$ 
for the initial state
$\ket{\downarrow,m_\ell=0}\otimes\ket{\nu^-,M=1}$ (purple line) is Lorentzian,
suggesting a criss-cross transition where both $m_\ell$ and $M$ change
by one. Indeed, we numerically find the transitions to
$\ket{\uparrow,m_\ell=-1}\otimes\ket{\nu^+,M=0}$ to be dominant.  In
contrast, around $\Vpm{92}$, 
the peak shows a deep dip due to a flip-flop transition to
$\ket{\uparrow,m_\ell=0}\otimes\ket{\nu^+,M=1}$.
Note that the cross-section does not vanish entirely at resonance because
our model includes all $m_\ell$/$M$-sub-levels. 
Repeating this analysis for all initial states in \figref{fig2}~\cite{appendix},
we find all resonances to be dominated by one type of transition. The electric
field controlled cross sections thus reveal whether a criss-cross or a flip-flop
transition  is at the core of a resonance.


\begin{figure}[t]
  \includegraphics{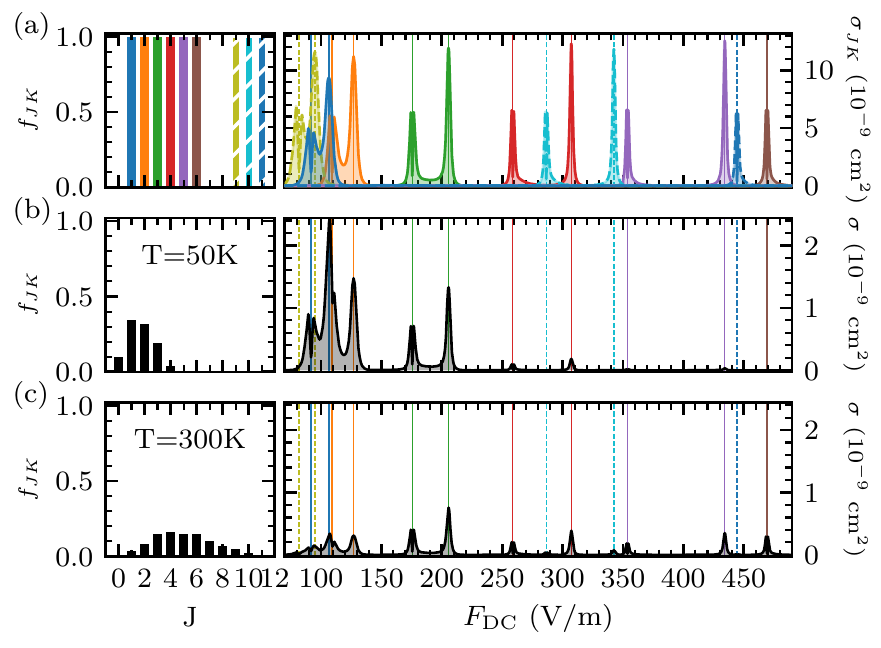}
  \caption{(Colour online)
    Rubidium Rydberg spectroscopy of ammonia with $M$-averaged cross sections
    shown on the right (for $v=10\,$m/s and $m_\ell=0$) and relative
    populations of the rotational states shown on the left for single
    rotational levels (a) and thermal ensembles (b,c) at rotational
    temperatures of $50$ and $300$~K, respectively.
    The vertical lines indicate resonances between the Rydberg transition and
    the inversion mode. Solid lines indicate $K=J$, dashed lines $K=J-1$
    (other molecular transitions are far off-resonant).
  }
  \label{fig4}
\end{figure}

\ssec{Molecules in an ensemble of rotational states}
We now discuss how to employ Rydberg spectroscopy to infer the population of
molecular states. \Figref{fig4} illustrates the working principle, correlating
rotational states with peaks in the cross section. The dependence of the
$M$-averaged cross section on the rotational quantum numbers $J$ and $K$, shown
in \figref[a]{fig4}, can be observed as a function of the DC field strength, due
to dependence of the inversion splitting on $J$ and $K$, cf.\
\myeqref{eq:w_inv}. \Figref{fig4} focusses on states with the Rydberg atom
initially in $m_\ell = 0$ (for $m_\ell = 1$ see \cite{appendix}) and molecules
in $J\le 11$ which have an inversion splitting near-resonant to the Rydberg
transition at the considered electric field strengths and temperatures.  The
cross sections  in \figref[a]{fig4} display a sequence of resonances, similarly
to \figref[b]{fig2}, with the position and height of the peaks depending on $K$,
in perfect agreement with the scaling of \myeqref{eq:scaling}. When considering
a thermal ensemble, the cross sections are obtained by averaging over all states
in the ensemble, cf. \figref[b,c]{fig4}. The occupation of the rotational
states is then given by $f_{JK}(T) = g/\mathcal N e^{-\frac{E_{JK}}{k_B T}}$
with $T$ the temperature, $\mathcal N$ a normalisation factor, and $g$ the
statistical weight of the state~\cite{appendix}. At very low temperatures, $T\le
10$\,K, the $J=K=1$ contribution dominates, leading to a similar pattern as
shown in \figref[b]{fig2}. At higher rotational temperatures
(\figref[b,c]{fig4}), an increasing number of rotational states contributes. For
a relative velocity of $v=\mps{10}$, as shown in \figref{fig4}, the resonances
are easily resolved and assigned to the different rotational states.  The
resonances for $J=1,2$ between $70$ and $\Vpm{140}$ are the hardest to resolve
but \figref{fig2} shows that this is possible until at least $\mps{50}$.
Provided the velocity is sufficiently low for the peaks to be resolved,
measurement of the cross section as a function of the electric field strength
allows for inferring the rotational state composition of the molecular ensemble
from the position and height of the peaks, for an example see \cite{appendix}.
In order to determine the composition of states with other $K$ values, simply
a different Rydberg transition has to be selected, e.g.\ $\ket{47p,m_\ell}$ to
$\ket{46d,m_\ell'}$.


\ssec{Conclusions}
We have derived the principles of collision-mediated, non-destructive
spectroscopy of polar molecules based on resonant dipole-dipole interaction with
Rydberg atoms, using a complete dynamical description of the collision. Taking
ammonia and rubidium as example, we  have shown that electric field control of
the cross sections allows for  inferring the relative population of rotational
states for velocities below  $\mps{100}$. The lineshape reveals the dominant
type of FRET-induced transition. For very low velocities, of the order of
$\mps{1}$, linewidths below 1$\,$MHz will be obtained. A key advantage of
Rydberg atom-enabled spectroscopy is the ability to measure spectra for
extremely low molecular density. Detection seem of single molecules in optical
tweezers or at molecular densities as low as $\unit[10^{5}]{cm^{-3}}$ seems
realistic. Our example of ammonia
and rubidium is easily carried over to other atomic and molecular species.
Exchanging rubidium by helium~\cite{Gawlas2020}, for instance, mainly reduces
the  quantum defects, rendering the isolation of a suitable dipole-dipole
transition from neighbouring (possibly higher-order) transitions somewhat more
difficult. When replacing ammonia by other polar molecules, purely rotational
transitions can be used instead of the inversion
mode~\cite{Zeppenfeld2017}. A method for state-resolved non-destructive
detection of polar molecules addresses an essential need for their application
in quantum technologies and cold reaction dynamics studies and establishes
Rydberg atoms as a versatile addition to the quantum control toolbox for cold
and ultracold molecules~\cite{Kuznetsova2018,Wang2022,Zhang2022}.



\begin{acknowledgments}
  We thank Ed Narevicius, Ronnie Kosloff, and Melanie Schnell for insightful
  discussions. Financial support from the Studienstiftung des deutschen Volkes
  e.V.\ and the Deutsche Forschungsgemeinschaft via the Priority Programme GiRyd
  (KO 2301/14-1 Grant No. 428456483) and Grant No. ZE 1096/2-1  is  gratefully
  acknowledged.
\end{acknowledgments}


\clearpage
\begin{appendix}

\section{Details on the molecular model}

  We consider FRET between Rydberg atoms and the inversion vibrational mode of
  ammonia. The latter can be modelled as a symmetric top rotor and its states
  can be described in terms of the symmetric top eigenstates $\ket{JKM}$.
  We use the standard notation with angular momentum quantum number $J$ and
  projections onto the molecular symmetry axis $K$ and onto the space-fixed
  $Z$-axis $M$~\cite{Kroto1975}. The inversion mode then couples the $+K$ and
  $-K$ states.
  The rotational subspace with $J=K=0$ does not possess an inversion mode and
  thus does not contribute to the FRET signal.
  At a given temperature, the occupation of the rotational states depends on $J$
  and $K$ and is given by $f_{JK}(T) = g/\mathcal N e^{-\frac{E_{JK}}{k_B T}}$
  with $T$ the temperature, $\mathcal N$ a normalization factor, and $g$ the
  statistical weight of the state. The statistical weight is $g=2(2J+1)$ except
  for $K=3m$ with $m\geq 1$ where $g=4(2J+1)$ due to the three-fold symmetry of
  ammonia~\cite{Townes1975}.

\section{Validity of the model}

  Our model relies on two basic assumptions --- the translational motion can be
  treated classically, and the state of the Rydberg atom faithfully represents
  the F\"orster resonant energy transfer with the polar molecule. In the
  following, we assess the validity of these assumptions.

  (1) The approximation of a classical, straight trajectory only remains valid
  as long as the kinetic energy $E_\text{kin}$ is much larger than the
  interaction strength $\coup$ between the particles. The latter depends on the
  distance between the two particles.
  A suitable value for this distance in this context is impact parameter at
  which the population exchange between the particles is maximal, which we term
  the critical impact parameter.  Its scaling with the relative velocity can be
  derived as follows.
  For the particles to exchange population, the phase accumulated by the
  particles, $\Phi = \coup T$, needs to be of the order of one, $\Phi \sim
  1$. The interaction scales as $\coup \sim r^{-3}$ (cf.\ \myeqref{eq6:dip_pot})
  and the time during which the two particles interact significantly can be
  approximated as $T \sim r/v$. Combining the equations and solving for the
  critical impact parameter $r=b^*$ results in the scaling
  \begin{align}
    b^* = \frac{c}{\sqrt{v}}.
    \label{seq:crit_b}
  \end{align}
  The proportionality constant $c$ is obtained from the numerical simulations
  by multiplying the impact parameter at which the exchange probability is
  maximal with the square-root of the velocity.
  We find the constant to be $c = \unit[1.66\cdot 10^{-6}]{m \sqrt{\frac{m}{s}}}$.
  For $\mps{0.1}$, the critical impact parameter amounts to $b^* = \microm{5}$.
  We find that the condition for the energy scale separation is fulfilled with
  the kinetic energy, $E_\text{kin} = \kHz{180}$, being three orders of
  magnitude larger than the interaction energy, $\coup = \kHz{1.6}$. Only at
  relative velocities below $\mps{8\cdot 10^{-6}}$ 
  will the two energies become equal ($\unit[0.001]{Hz}$).

  (2) Furthermore, assuming a classical trajectory also relies on the de
  Broglie wavelength ($\lambda_B = h/\mu v$ with reduced mass $\mu$) being
  much smaller than the distance between the particles.
  For $\mps{0.1}$, the critical impact parameter, $b^* = \microm{5}$,
  is one order of magnitude larger than the de Broglie wavelength,
  $\lambda_B = \microm{0.3}$.
  Only at relative velocities below $\mps{3\cdot 10^{-4}}$ 
  will the de Broglie wavelength become equal to the critical impact parameter
  ($\microm{100}$). 


  (3) The duration of the experiment is limited by the lifetime of the Rydberg
  atom. For the Rydberg states considered in this work and at room temperature,
  the lifetime is mainly limited by decay processes induced by blackbody
  radiation and is around $\microsec{75}$ (calculated using the \emph{ARC}
  library~\cite{Sibalic2017}).
  The interaction time which is necessary for the particles to exchange
  a significant amount of population is easiest approximated when assuming a
  quasi-stationary setup in which the beginning and end of the experiment is
  defined by the excitation of the atom to the Rydberg regime and its
  ionisation. Using the critical impact parameter as derived above, the
  interaction time can be approximated as $T = \frac{2 b^*}{v}$. Inserting
  \myeqref{seq:crit_b}, solving for $v$ and inserting as interaction time the
  lifetime of the Rydberg atom, we obtain the critical velocity
  of $\mps{0.1}$. 
  As a result, the model is mainly limited by the lifetime of the Rydberg atom
  and remains valid approximately until relative velocities of $\mps{0.1}$.

\section{Approximation of necessary molecular densities}

  In the following, we elaborate on the estimation of the molecular density which
  required to detect the molecules via their interaction with Rydberg atoms.
  First, the effective volume that a Rydberg atom probes during a given time
  $T$ can be approximated as $V = T v \sigma$. When inserting the peak values of
  the cross section from \figref{fig2}(b-d), for instance $\sigma \sim
  \cmtwo{2\cdot 10^{-7}}$ at $v = \mps{1}$, and assuming an interaction time of
  $T = \microsec{100}$, we obtain the value of $V = \unit[2 \cdot
  10^{-9}]{cm^3}$ as given in the main paper. This value is independent of $v$
  as $\sigma$ scales as $1/v$, cf. \myeqref{eq:scaling}. Second, at a molecular
  density $\varrho$, the Rydberg atom interacts on average with $N = \varrho V$
  molecules. At a molecular density of $\unit[5 \cdot
  10^{8}]{cm^{-3}}$, the Rydberg atom then interacts on average with exactly one
  molecule and the signal is saturated. At higher densities, our approximation
  of a dilute medium breaks down. At lower densities, the signal reduces
  accordingly.  When trapping, for instance, a single molecule in an optical
  tweezer with an estimated volume of $\unit[10^{-8}]{cm^{3}}$, this corresponds
  to a density of $\unit[10^{8}]{cm^{-3}}$ and $20\%$ of the saturated signal is
  achievable.

  In a real experiment, these values have to be compared to the measured
  background. In the experiment~\cite{Jarisch2018}, the background is dominated
  by blackbody radiation which causes the Rydberg atom to change its state even
  if no molecules are present. During an interaction time of $\microsec{100}$,
  $2\%$ of the Rydberg atoms can be found in $45d$ due to this effect.
  Therefore, at a molecular density of $\unit[10^{7}]{cm^{-3}}$, the FRET signal
  and the background are of the same order of magnitude. This value can be
  further reduced by a few orders of magnitude: at $\unit[10^{5}]{cm^{-3}}$, the
  FRET signal is hundred times smaller than the background but observing
  a change of $1\%$ in in the background transfer rate seems realistic.
  Of course, going to a setup with shielding of room temperature blackbody
  radiation, i.e., a cryogenic setup, would reduce the lower bound on the
  density for which molecules can be detected even further.

\section{Detailed analysis of the peak structure in the electric field
controlled cross section of Fig. 2}

  \Figref{fig2} shows an intricate pattern of dips and peaks in the cross
  section as a function of the electric field. As discussed in the main text,
  the line shape can be traced back to the dominant transition occurring in the
  system. The possible kind of transitions, as discussed with the help of
  \figref{fig3} in the main text, are summarised in Tab.~\ref{tab:transitions}.
  \begin{table}[b]
    \centering
    \begin{ruledtabular}
    \begin{tabular}{c c c}
      name & selection rule & line shape \\
      \hline
      criss-cross        & $\Delta m_\ell = \Delta M = \pm 1$ & peak \\
      linear flip-flop   & $\Delta m_\ell = \Delta M =     0$ & dip \\
      diagonal flip-flop & $\Delta m_\ell =-\Delta M = \pm 1$ & dip \\
    \end{tabular}
    \end{ruledtabular}
    \caption{%
      Overview of the types of transitions caused by dipole-dipole interaction.
    }
    \label{tab:transitions}
  \end{table}
  This, together with the location of the resonance allows one to determine the
  dominant transition occurring in the combined system of Rydberg atom and
  molecule. (1) The peak position of resonance as a function of electric field
  strength reveals which
  transition occurs in the Rydberg atom as shown in \figref[a]{fig2}. This gives
  the value of $\Delta m_\ell$. (2) The line shape reveals whether a criss-cross
  or flip-flop transition occurs as shown in Tab.~\ref{tab:transitions}. (3)
  Combining both, the value of $\Delta M$ can be concluded. We conduct this
  procedure in the following for all lines shown in \figref{fig2}.

  \begin{figure*}
    \begin{overpic}{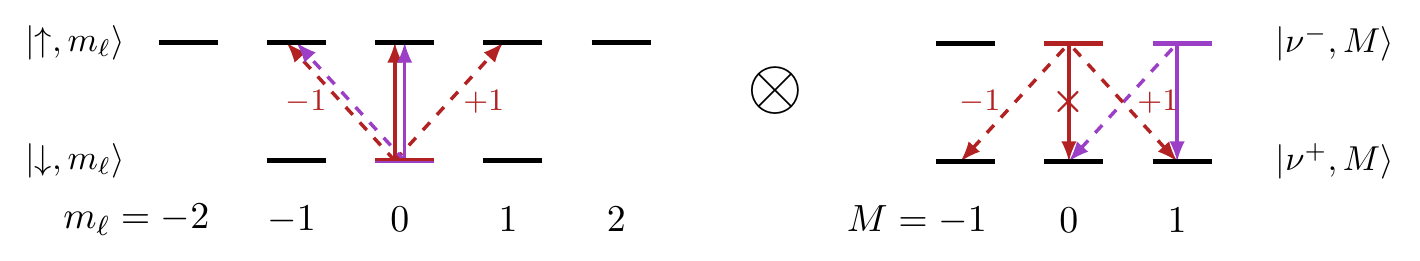}
      \put (-8,15) {(a)}
    \end{overpic}
    \begin{overpic}{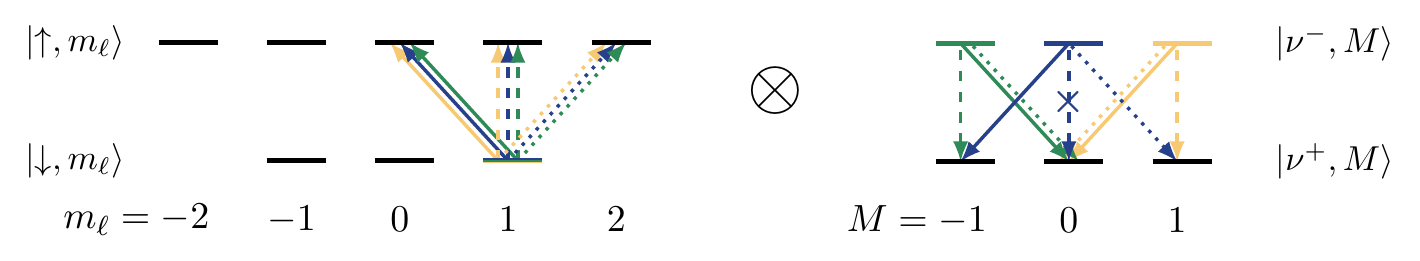}
      \put (-8,15) {(b)}
    \end{overpic}
    \caption{Dominant transition in the combined system of Rydberg atom (left)
    and molecule (right) as shown in \figref{fig2}.
    (a) shows transitions for initial states with $m_\ell = 0$.
    Solid lines indicate the resonance \res{0}{0} which occurs at $\Vpm{92}$,
    dashed lines indicate the resonance \res{0}{\pm 1} which occurs at $\Vpm{107}$.
    (b) shows transitions for initial states with $m_\ell = 1$.
    Solid lines indicate the resonance \res{1}{0} which occurs at $\Vpm{77}$,
    dashed lines indicate the resonance \res{1}{1} which occurs at $\Vpm{85}$,
    dotted lines indicate the resonance \res{1}{2} which occurs at $\Vpm{149}$.
    The line colours of the arrows indicate the initial state using the same
    colour code as in \figref[b-d]{fig2}.
    }
    \label{figs1}
  \end{figure*}

  In the main text, we have discussed as an example the initial state
  $\initprodstate{0}{1}$ (purple line in \figref[b-d]{fig2}). The dominant
  transitions are furthermore illustrated in \figref[a]{figs1}. We found the
  peak at $\Vpm{107}$ (corresponding to the resonance \res{0}{\pm 1}) being due
  to a criss-cross transition to
  $\ket{\uparrow,m_\ell=-1}\otimes\ket{\nu^+,M=0}$ (purple dashed arrows in
  \figref{figs1}). The dip around $\Vpm{92}$ (\res{0}{0}), on the other hand,
  can be attributed to a linear flip-flop transition to $\finlprodstate{0}{1}$
  (purple solid arrows in \figref{figs1}).

  The initial state $\initprodstate{0}{0}$ (red) shows a qualitatively similar
  behaviour as $\initprodstate{0}{1}$ (purple), since the Rydberg atom is
  initially in the same state. At $\Vpm{107}$, criss-cross transitions with both
  $\Delta m_\ell = \Delta M = \pm 1$ are allowed (as indicated next to the red
  dashed arrows in \figref[a]{figs1}). Moreover, the linear flip-flop
  transition to $\finlprodstate{0}{0}$, suggested by the dip at $\Vpm{92}$
  (\res{0}{0}), is forbidden by the molecular selection rule $M=0
  \not\leftrightarrow M'=0$ (as indicated by the red solid arrows in
  \figref[a]{figs1}). Instead, the molecule performs transitions with $\Delta
  M = \pm 1$ leading to a dip in the electric field controlled cross section.

  The other three initial states correspond to the Rydberg atom being initially
  in $m_\ell = 1$ and thus relate to the resonances indicated by vertical blue
  lines in \figref{fig2}. The dominant transitions are illustrated in
  \figref[b]{figs1}.
  We continue with the initial state $\initprodstate{1}{0}$ (blue). The cross
  section of this initial state shows a strong peak at the $\res{\pm 1}{\pm 2}$
  transition at $\Vpm{149}$. The line shape clearly indicates a criss-cross
  transition (i.e.\ $\Delta m_\ell = \Delta M = \pm 1$).  The sign of the
  criss-cross transition can be deduced from the resonance itself which reveals
  that the Rydberg atom performs a transition from $46p,m_\ell = 1$ to
  $45d,m_\ell = 2$. Combining the two insights, we can deduce that the dominant
  transition occurs to the state $\finlprodstate{2}{1}$ (dotted blue lines in
  \figref[b]{figs1}).
  The cross section also forms a peak around the resonance at $\Vpm{77}$
  (\res{\pm 1}{0}). Using similar arguments, the dominant transition is also
  a criss-cross transition to the state $\finlprodstate{0}{-1}$ (solid blue
  lines in \figref[b]{figs1}).
  Around the resonance at $\Vpm{85}$ (\res{\pm 1}{\pm 1}), the cross section
  forms a dip indicating a flip-flop transition. As the resonance indicates the
  Rydberg atom to perform a transition from $46p,m_\ell = 1$ to $45d,m_\ell
  = 1$, we can identify the linear flip-flop transition to
  $\finlprodstate{1}{0}$ to be dominant (dashed blue lines in
  \figref[b]{figs1}). However, this transition is again forbidden by the
  molecular selection rule $M=0 \not\leftrightarrow M'=0$. Instead, the molecule
  performs transitions with $\Delta M = \pm 1$ which equally leads to a dip in
  the electric field controlled cross section.

  The initial state $\initprodstate{1}{-1}$ (green) shows a qualitatively very
  similar behaviour to $\initprodstate{1}{0}$ (blue), since similar transitions
  are involved. The only qualitative difference between the two occurs at the
  resonance around $\Vpm{77}$ (\res{\pm 1}{0}). The green line forms dip here
  while the blue one forms a peak. The green state therefore performs a diagonal
  flip-flop transition to $\finlprodstate{0}{0}$ (solid green lines in
  \figref[b]{figs1}).

  Finally, the state $\initprodstate{1}{1}$ (yellow) also shows a similar
  behaviour to $\initprodstate{1}{0}$ (blue), but deviates around the resonance
  at $\Vpm{149}$ (\res{\pm 1}{\pm 2}) as it shows a dip instead of a peak.
  Instead of a criss-cross, this state performs a diagonal flip-flop transition
  with $\Delta m_\ell = -\Delta M = 1$, driving the population to
  $\finlprodstate{2}{0}$ (dotted yellow lines in \figref[b]{figs1}) and causing
  the strong dip in the spectrum.

  It shall lastly be noted that cross sections do not completely vanish at
  resonance for flip-flop transitions (which form a dip) in the realistic model
  shown in \figref{fig2}, while in the simplified model of \figref{fig3} they
  do. The reason is that for multi-level systems, the dynamics cannot be
  limited to a single transition. For instance, we have discussed that the
  initial state $\initprodstate{1}{1}$ (yellow state in \figref{fig2}) leads to
  a dip in the cross section at the \res{\pm 1}{\pm 2} transition at $\Vpm{149}$
  thus transferring population into $\finlprodstate{2}{0}$. This
  state could further be transferred to $\initprodstate{1}{-1}$ via
  a criss-cross transition.  Thus, a fraction of the population is now in the
  state which we indicated by the green colour. In a second order process, this
  state can thus undergo a criss-cross transition leading to a peak at the
  considered $\res{\pm 1}{\pm 2}$ transition which overlays with the initial dip
  created by the dominant flip-flip transition.

\section{Molecules in an ensemble of rotational states for $m_\ell = 1$}

  \begin{figure*}
    \includegraphics{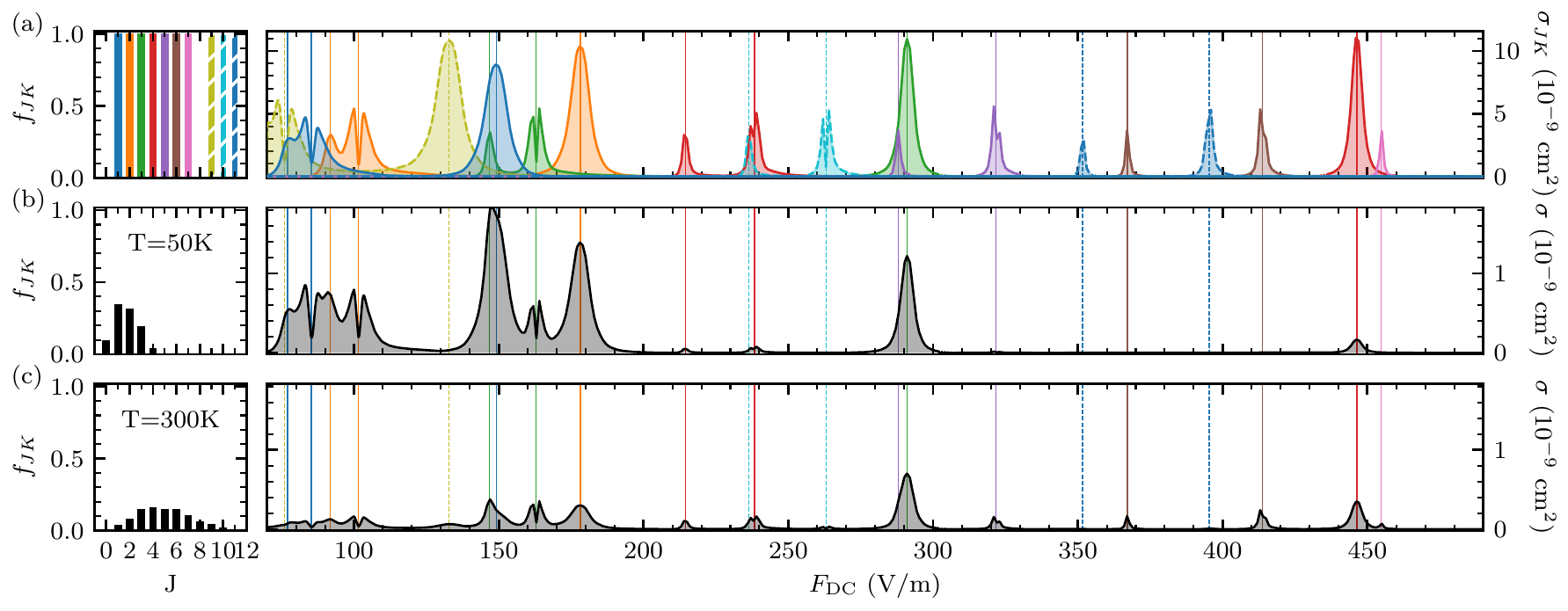}
    \caption{Like \figref{fig4} but for $m_\ell = 1$:
      Rubidium Rydberg spectroscopy of ammonia with $M$-averaged cross sections
      shown on the right (for $v=10\,$m/s and $m_\ell=1$) and relative
      populations of the rotational states shown on the left for single
      rotational levels (a) and thermal ensembles (b,c) at rotational
      temperatures of $50$ and $300$~K, respectively.
      The vertical lines indicate resonances between the Rydberg transition and
      the inversion mode. Solid lines indicate $K=J$, dashed lines $K=J-1$
      (other molecular transitions are far off-resonant).
    }
    \label{figs2}
  \end{figure*}

  We now present the dependence of the electric field controlled cross section on
  the rotational quantum numbers $J$ and $K$ assuming the Rydberg atom to be
  initially in $46P,m_\ell = 1$. This is equivalent to the results presented in
  the section ``Molecules in an ensemble of rotational states'' of the main
  paper but changing $m_\ell$ from $0$ to $1$. The results are shown in
  \figref{figs2} where three resonances appear for each rotational state
  $\ket{J,K}$,
  giving rise to a richer spectrum as compared to $m_\ell = 0$.
  It can be seen that, different from \figref{fig4}, several peaks overlap, such
  as the green ($J=K=3$) and the purple ($J=K=5$) one around $\Vpm{290}$. However,
  this does not hamper the fitting procedure. Each rotational sublevel gives rise
  to as least one peak which is sufficiently isolated from the others in order to
  determine its contribution to the cross section. For instance, even the $J=K=1$
  state (blue) can be clearly identified at $50$\,K by its characteristic
  double-peak structure close to $\Vpm{85}$ in \figref[b]{figs2}.

\section{Fitting the relative populations from measured cross sections}

  Measuring the electric field controlled cross section in an experiment allows
  for inferring the relative populations of rotational states. In the following,
  we sketch an example.
  First, we generate a noisy Boltzmann distribution which simulates an unknown
  distribution of rotational states in an experiment. We start from a
  Boltzmann distribution at $300$\,K (gray shade in left-hand panel of
  \figref[b]{figs3}) and add noise (red shade) by
  multiplying each component with a random number between $0$ and $2$, where
  a factor between $0$ and $1$ entails reduction and
  a factor between $1$ and $2$ increase of the corresponding component.
  We then re-normalise the distribution. When averaging the cross sections of
  single rotational levels (cf.\ \figref[a]{figs3}) accordingly, we obtain the
  signal shown in \figref[b, red line]{figs3}. This curve will serve as the
  unknown signal acquired in an experiment and will be the starting point for
  the fitting procedure. Note that we did not add any further disturbance to the
  signal to simulate experimental noise. In particular, we did not alter the
  peak positions or their shape but kept them as given in \figref[a]{figs3}.
  We expect the theoretical prediction to be very accurate, since the
  theoretical model of the Rydberg atom and the rotation and inversion
  mode of the molecule are very well known.

  \begin{figure*}
    \includegraphics{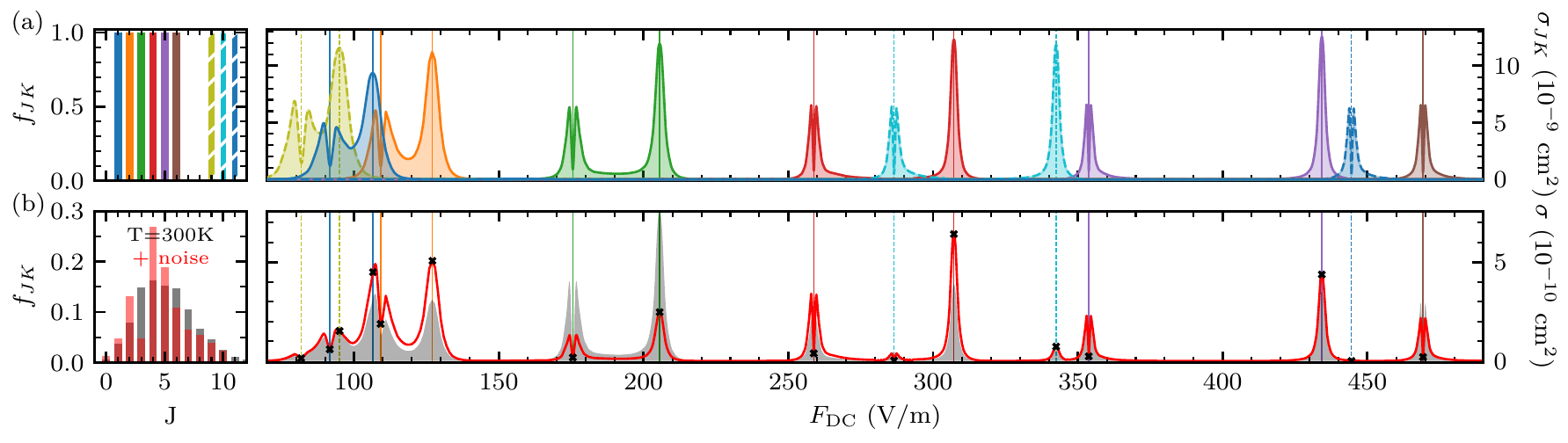}
    \caption{Similar to \figref{fig4}:
      Rubidium Rydberg spectroscopy of ammonia for $v=10\,$m/s and $m_\ell=0$.
      State-resolved cross sections are shown in (a), and (noisy) thermally
      averaged cross section in (b) panel. The red bar chart (b, left) and the
      red line (b, right) indicate the population distribution and corresponding
      cross sections when adding noise to a Boltzmann distribution with $300$\,K
      (black). The black crosses indicate the values used for fitting.
    }
    \label{figs3}
  \end{figure*}

  We test two different strategies for acquiring the composition of rotational
  states by fitting the theoretical data from \figref[a]{figs3} to our test
  signal from \figref[b, red]{figs3}. First, we use the full information on the
  state-dependent cross sections $\sigma_{JK}(F_\text{DC})$ as shown in
  \figref[a]{figs3} which have a very high resolution. The averaged cross
  section $\sigma(F_\text{DC})$ can be written as
  \begin{align}
    \sigma(F_\text{DC}) = \sum_{JK} \, c_{JK} \, \sigma_{JK}(F_\text{DC})
  \end{align}
  with $\sum_{JK} c_{JK} = 1$. We will treat the $c_{JK}$ as fitting parameters
  which directly reflect the relative population of rotational states. Due to
  the resonance conditions, only $9$ fitting parameters are left: $c_{11}$ to
  $c_{66}$ and $c_{98}$ to $c_{11\,10}$. We perform the fit using the
  \emph{optimize} package of $\emph{scipy}$ with guess parameters $c_{JK} = 1$.
  The guess signal is therefore identical to the sum on the peaks in
  \figref[a]{figs3}. We find that the fit recovers the input parameters
  exactly and
  the fitted curve lies exactly on top of the red line in \figref[b]{figs3}.

  To test the applicability of the fitting procedure to real experimental data,
  we reduce the resolution of the signal. Namely, we only consider data points
  which are located directly at the resonances as indicated by the black crosses
  in \figref[b]{figs3}, thus reducing the number of data points to $16$. Note
  that the signal is very low at some resonances because the line forms a dip
  around them. We repeat the fitting procedure with this decreased
  resolution and find the correct input parameters with a relative error of
  $10^{-16}$.
  Note that, when dealing with very small signals in a real experimental
  setting, it might be beneficial to measure the cross sections slightly next to
  the resonance. This increases the signal-to-noise ratio when detecting cross
  sections which form a dip at the resonance.

  This example demonstrates that the relative populations of rotational states
  can be inferred with a very high resolution from a given input signal. Of
  course, the error of the fitting procedure will ultimately be given by the
  error bars of the experiment.
\end{appendix}

\end{document}